\documentclass[prb,aps,reprint,twocolumn]{revtex4}

\usepackage{graphicx}
\usepackage{amsmath}
\usepackage{amsfonts}
\usepackage{amssymb}
\usepackage{upgreek}
\usepackage{epstopdf}
\usepackage[usenames,dvipsnames]{xcolor}
\usepackage{ulem}

\renewcommand{\imath}{i}

\newcommand{\TS}{T_{\mathrm{sim}}}

\begin{document}

\title{Particle number scaling for diffusion-induced dissipation in graphene and carbon nanotube nanomechanical resonators}

\author{Christin Rh\'en}
\email{christin.rhen@chalmers.se}
\author{Andreas Isacsson}
\email{andreas.isacsson@chalmers.se}

\affiliation{Department of Physics\\
Chalmers University of Technology\\
SE-412 96 G\"oteborg\\
Sweden}

\date{\today}

\begin{abstract} 
When a contaminant diffuses on the surface of a nanomechanical resonator, the motions of the two become correlated. Despite being a high-order effect in the resonator-particle coupling, such correlations affect the system dynamics by inducing dissipation of the resonator energy. Here, we consider this diffusion-induced dissipation in the cases of multiple particles adsorbed on carbon nanotube and graphene resonators. By solving the stochastic equations of motion, we simulate the ringdown of the resonator, in order to determine the resonator energy decay rate. We find two different scalings with the number of adsorbed particles $K$ and particle mass $m$. In the regime where the adsorbates are inertially trapped at an antinode of vibration, the dissipation rate $\Gamma$ scales with the total adsorbed mass $\Gamma\propto Km$. 
In contrast, in the regime where particles diffuse freely over the resonator, the dissipation rate scales as the product of the total adsorbed mass and the individual particle mass: $\Gamma\propto Km^2$. 
\end{abstract}

\maketitle


\section{Introduction}
\begin{figure}[t]
\centering
\includegraphics[width=\linewidth]{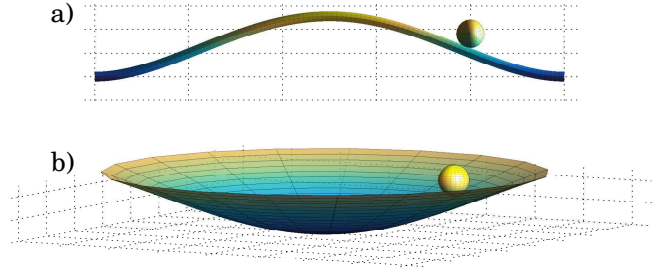}
\caption{(color online) A particle adsorbed on the surface of a suspended a) carbon nanotube beam resonator b) graphene drum resonator. Thermal fluctuations cause the adsorbate to perform a random walk over the resonator surface, which induces dissipation in the vibrational motion. 
Here, only a single particle is shown for clarity, whereas in the following we consider up to several hundred non-interacting adsorbates.
\label{fig:system}}
\end{figure} 
Nanoelectromechanical resonators are important components in today's state-of-the-art sensors of charge~\cite{Lassagne_2009, Steele_2009,Meerwaldt_2012, Hakkinen_2015}, spin~\cite{Rugar_2004}, force~\cite{Mamin_2001,Moser_2013}, position~\cite{Knobel_2003}, and mass~\cite{Bachtold_2008, Zettl_2008, Atalaya_2010, Roukes_2011, Eom_2011, Roukes_2012, Chaste_2012,Tavernarakis_2014}. In particular, resonators fabricated from suspended carbon nanotubes (CNTs) have attracted significant interest, since they combine an extremely low mass with a high mechanical stiffness~\cite{Yakobson_2001, Yang_2011}: a combination that enables unprecedented sensitivity.

A related but less mature technology is based on suspended graphene membranes~\cite{Hone, Hakonen, McEuen, Desmukh, Hone_VCO}. These share many of the outstanding material properties of CNT resonators, in addition to allowing top-down fabrication\cite{Zande} and having a higher interaction cross-section with particles. However, graphene resonators suffer from higher masses and lower resonant frequencies than their CNT counterparts. Despite this, mechanical quality factors \mbox{(Q-factors)} of $10^5$ have been reported~\cite{Eichler_2011}, approaching the highest observed Q-factors of CNT resonators~\cite{Moser_2014}. 

In general, resonator-based sensing schemes rely on detecting minute shifts in the resonant frequency. Hence, in the pursuit of increased sensitivity, significant attention has been devoted to understanding the microscopic origins of spectral broadening in nanoscale resonators. There are two types of such broadening, elastic and inelastic, caused by different types of processes. Elastic broadening, dephasing, arises from uncorrelated frequency fluctuations, whereas inelastic broadening is related to the dissipation of vibrational energy. Dephasing can occur, for instance, due to thermoelastic fluctations~\cite{McEuen_PNAS}, strong electromechanical coupling~\cite{Micchi_2015} or, in the case of mass sensors, due to fluctuating mass loads~\cite{Dykman_2010, Atalaya_2011, Atalaya_2011_2, Atalaya_2012,Roukes_Diffusion}. To distinguish dephasing from dissipation, time domain techniques, such as ring-down measurements, have recently been developed~\cite{Leeuwen_2014,Schneider_2014}.

In addition to resonator dephasing from diffusing adsorbates, extensively studied in  Ref.~[\onlinecite{Atalaya_2011}] for a single particle and Ref.~[\onlinecite{Atalaya_2012}] for several particles, inertial back-action on the adsorbates causes the frequency noise to become correlated with the resonator motion~\cite{Edblom_2014}. The existence of such correlations can lead to parametric attenuation of the resonator. To capture this effect, the system must be modelled beyond the usual rotating wave approximation. Diffusion-induced dissipation is thus a higher-order effect. In the single-particle case, as we showed in Ref.~[\onlinecite{Edblom_2014}], this dissipation is expected to be small. However, a more realistic scenario is that several adsorbates will be present; we here consider the question of how the dissipation rates change under such circumstances.

Studying a 1D resonator with a single adsorbed particle (Fig.~\ref{fig:system}a), we showed~\cite{Edblom_2014} that the mechanical energy dissipation displays two characteristic behaviors. For high oscillation amplitudes, the particle becomes trapped near the antinode of vibration due to inertial forces, whereas for low amplitudes the adsorbate can diffuse freely across the resonator. In the former scenario, the {\it trapping regime}, the energy decay is linear in time, while in the the latter, the {\it diffusive regime}, it is that of a nonlinearly damped resonator. 

\begin{figure}[t]
\centering
\includegraphics[width=\linewidth]{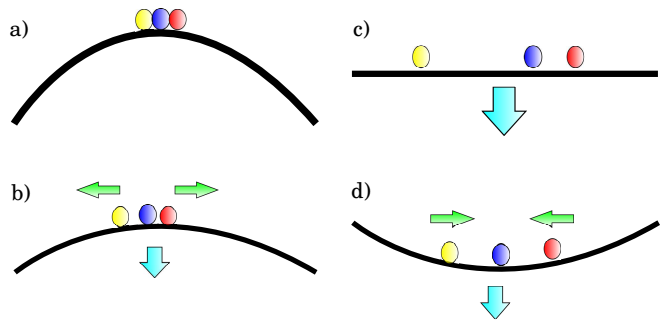}
\caption{(color online) Illustration of the short timescale-dynamics of a resonator-particle system, in the trapped regime. a)~The resonator is at its maximum positive deflection, and the particles are inertially trapped at the antinode. b)~As the resonator amplitude decreases, so does the strength of the inertial trapping, and thermal fluctuations in the particle positions are no longer suppressed. c)~The resonator is relaxed and the particles diffuse freely. The typical distance covered by each particle is determined by the diffusion constant $D\propto k_{\mathrm B}T$. d)~As its amplitude once more increases, the resonator does work on the particles as the inertial trap reappears. At the maximum negative resonator deflection, the adsorbates are once more trapped at the antinode, and the cycle begins anew. The net result is a periodic force on the adsorbates, with twice the frequency of the resonator motion.\\
Note that the inertial force exerted on a particle at position $x$ is proportional to $\ddot w|_x\nabla w|_x$, i.e., the product of the resonator acceleration and slope. Because of this, the inertial potential is identical in panels b) and d).}
\label{fig:dynamics}
\end{figure}

In the trapping regime the maximum mechanical energy lost per cycle is of the order of $\Delta E_{\rm m}\sim k_{\rm B}T$. This maximal damping is achieved when the resonator performs the maximum amount of work  each cycle as it restores the particle to an inertially trapped position at the antinode of vibration, by working against the thermal fluctations. This situation is schematically depicted in Fig.~\ref{fig:dynamics}. However, for such a situation to arise, the inertial trapping must be strong, requiring a large mechanical energy $E_{\rm m}$. Consequently, the ``Q-factor'' $E_{\rm m}/\Delta E_{\rm m}$ is always large in the case of a single adsorbate.

In this paper, we further investigate the interplay between resonators and Brownian adsorbates, by modelling the ringdown of one- and two-dimensional carbon nanoresonators on which several particles are adsorbed (see Fig.~\ref{fig:system}). The number of particles, here denoted by $K$, ranges from $K=1$ to $K\gtrsim10^2$. We find (see Sec.~\ref{subsec:Trapregime_single_mode}) that in the inertial trapping regime, the dissipation rate increases linearly with the number of particles. Thus, the maximal damping rate can now reach $\Delta E_{\rm m}\sim K k_{\rm B}T$. For adsorbates having individual masses $m$ this leads to an energy decay rate $\Gamma\propto Km=m_{\rm tot}$. 

In contrast to the trapped dynamics, we find for the diffusive regime (see Sec.~\ref{subsec:Diffregime_single_mode}) that while the dissipation rate still scales linearly with the total number of added particles $K$, it is no longer only dependent on the total added mass. Instead, it scales as $\Gamma\propto Km^2=m_{\mathrm{tot}}m$. Hence, when the particles can diffuse over the resonator their mass distribution will play a role for the system dynamics.

In both regimes, the adsorbates induce a mode coupling between flexural eigenmodes, ensuring that the system eventually thermalizes. In the inertial trapping regime, it has been shown~\cite{Edblom_2014} that this mode coupling enhances the fundamental flexural mode dissipation rate through intermode energy exchange. Here, we study the effect of the mode coupling in the diffusive regime. We find (see Sec.~\ref{subsec:Diffregime_many_modes}) that while the mode coupling still leads to a rapid thermalization of the higher modes, the ringdown time of the initially excited fundamental mode is largely unaffected.

\section{Equations of motion \label{sec:EOM}}
We consider a system consisting of a carbon resonator of mass $M$ on which $K$ particles are adsorbed. The resonator is either a doubly clamped carbon nanotube, Fig.~\ref{fig:system}a), or a circular graphene sheet pinned along the edges, Fig.~\ref{fig:system}b), and is initially excited in its fundamental flexural vibration mode. The subsequent free evolution of the resonator-adsorbate system is studied by deriving and solving its stochastic equations of motions. In general, any vibrational mode can be initially excited, e.g. by matching the actuating voltage frequency to that of the chosen mode.

Considering only flexural vibrations, the Lagrangian density for a resonator of mass density $\rho$ is~\cite{LL}
\begin{equation}
\mathcal L_0=\tfrac12\rho \dot w^2-\tfrac12\sigma[\nabla w]^2-\tfrac12\upkappa[\nabla^2w]^2.
\end{equation}
Here, $w(X,t)$ is the transverse resonator displacement at position $X$ along the resonator surface. The tension $\sigma$ is assumed to be constant: a good approximation in the limit of small vibrations or large prestrain. Finally, we note that for an atomically thin graphene membrane, the bending rigidity $\upkappa$ can be neglected~\cite{Lindahl_2012}. 

Including the $K$ non-interacting particles with masses $m_\kappa$, adsorbed at positions $x_\kappa$, the full Lagrangian density is found to be
\begin{equation}
\mathcal L=\mathcal L_0+\tfrac12\sum_{\kappa=1}^Km_\kappa\delta(x_\kappa-X)\left[\dot x_\kappa^2+(\dot w+\dot x_\kappa \cdot\nabla w)^2\right].
\end{equation}
Each term in the sum is the kinetic energy $\frac12m_\kappa\dot{\mathbf r}_\kappa^2$ of the $\kappa$:th particle. For simplicity, we consider identical adsorbates: $m_\kappa\equiv m$. Relaxing this constraint is a straightforward generalization of the present work.
 
By expanding the displacement field in eigenmodes, $w(X,t)=\sum_nq_n(t)\phi_n(X)$, and varying the Lagrangian, we arrive at the equations of motion
\begin{align}
\ddot q_n+\omega_n^2q_n-\epsilon\sum_{\kappa,\ell}\phi_n(x_\kappa)\phi_\ell(x_\kappa)\omega_\ell^2q_\ell&=0,\label{eq:eom1}\\
\dot x_\kappa-{\gamma}^{-1}\sum_{j,\ell}\omega_j^2q_jq_l\phi_j(x_\kappa)\phi'_\ell(x_\kappa)&=\sqrt D\eta_\kappa(t).
\label{eq:eom2}
\end{align}
For details of the mode shapes $\phi_n(X)$ and frequencies $\omega_n$, see Refs.~[\onlinecite{Edblom_2014}] and~[\onlinecite{Eriksson_2013}]. 

Upon deriving Eq.~(\ref{eq:eom1}) from the Lagrangian, inertial forces of the type $\mathrm d^2/\mathrm dt^2[w(x_\kappa(t),t)]$ appear. These terms can be simplified using that the equation of motion is here derived under the assumption that $(q_n/L)^2(\omega_n/\gamma)\ll 1$; the resonator vibration amplitude is small compared to the characteristic system size $L$. Then, we may replace the total time derivative with the partial derivative: $\mathrm d^2/\mathrm dt^2[w(x_\kappa(t),t)]\rightarrow \partial^2/\partial t^2 [w(x_\kappa(t),t)]$. Working to lowest order in $\epsilon=m/M$, it is further permissible to make the approximation $\epsilon \partial_t^2 w(x_\kappa(t),t)\approx-\epsilon\sum_j\omega_j^2q_j(t)\phi_j(x_\kappa)$, yielding the coupling term in Eq.~\eqref{eq:eom1}.

In order to model thermal fluctuations in the adsorbate positions, a stochastic force $\eta_\kappa$ has been introduced in Eq.~\eqref{eq:eom2}. We let $\eta_\kappa$ be $\delta$-correlated, \mbox{$\langle\eta_\kappa(t)\eta_{\kappa'}(t')\rangle=\delta_{\kappa,\kappa'}\delta(t-t')$}, and denote the associated damping rate ${\gamma}$. The diffusion constant is \mbox{$D=2k_{\rm B}T/m{\gamma}$}. By considering an overdamped regime, $m\ddot x_\kappa\ll m{\gamma}\dot x_\kappa$, the inertial term has been neglected.

The number of adsorbed particles that can be modelled is limited by the computational resources required to numerically integrate the system of equations~\eqref{eq:eom1}-\eqref{eq:eom2}. In the present study, we consider $K\leq 400$; our results are straightforwardly extrapolated to larger $K$ as long as the total relative added mass is small, $K\epsilon\ll1$, and the assumption of non-interacting particles holds. 

\subsection{Single-mode approximation}
We have previously~\cite{Edblom_2014} seen that the difference between the full multimode ring-down dynamics of a resonator-particle system and the approximative single-mode case is quantitative, not qualitative. Hence, we initially let the resonator motion include only its fundamental vibrational mode, and shift to action-angle variables $\mathcal E(t),\theta(t)$. These are defined through
\begin{gather}
q_0=L\sqrt{\mathcal E}\cos(\omega_0t+\theta)\nonumber\\
\dot q_0=-\omega_0L\sqrt{\mathcal E}\sin(\omega_0t+\theta)
\end{gather}
where $L$ is the length (radius) of the CNT (graphene) resonator. With $\nu=\omega_0 t+\theta$, the equations of motion become
\begin{align}
&\partial_\tau\mathcal E=-\epsilon\ \mathcal E\sin2\nu\sum_{\kappa}\phi_0^2(x_\kappa)\label{eq:actang1}\\
&\partial_\tau\theta=-\tfrac12\epsilon\left(1+\cos2\nu\right)\sum_{\kappa}\phi_0^2(x_\kappa)\\
&\partial_\tau x_\kappa=\tfrac12\gamma^{-1}\mathcal E(1+\cos2\nu)\partial_x\phi_0^2|_{x_\kappa}+\sqrt{\mathcal D}\eta_\kappa(\tau).
\label{eq:actang3}
\end{align}
In these equations, we have redefined $x_\kappa/L\rightarrow x_\kappa$ and $\gamma/\omega_0\rightarrow\gamma$ in order to eliminate dimensions. Additionally, $\tau = \omega_0t$ and $\mathcal D = 2 \TS/\epsilon\gamma$, where $\gamma$ is dimensionless and $\TS=k_{\rm B}T/ML^2\omega_0^2$.

\section{Results \label{sec:Results}}

\subsection{Many adsorbates in the inertial trapping regime\label{subsec:Trapregime_single_mode}}

To begin with, we consider the inertial trapping regime, in which the adsorbed particles fluctuate around the antinode of the fundamental mode vibration. This is the relevant regime if the initial resonator amplitude is high enough that the kinetic energy of a particle due to the vibrational motion of its substrate is much larger than the thermal energy: \mbox{$\mathcal E_0\equiv\mathcal E(\tau=0)\gg\gamma\mathcal D$}. Characteristic for this regime is a linear decay of energy $\mathcal E$ in time~\cite{Edblom_2014}.

We choose a coordinate system such that the fundamental mode antinode is located at $X=0$, and expand the mode function in Eq.~\eqref{eq:actang3}:
\begin{equation}
\partial_x\phi_0^2|_{x_\kappa}\approx 2\phi_0(0)\phi_0''(0)x_\kappa\equiv-kx_\kappa.
\end{equation}
Equation~\eqref{eq:actang3} then becomes linear and can be formally solved. Its solution is substituted back into equation~\eqref{eq:actang1}, which is subsequently averaged over fast oscillations (for details, see Ref.~[\onlinecite{Edblom_2014}]). Since the particles are identical, so is the contribution from each $\kappa$ to the sum in~\eqref{eq:actang1}, leading to a factor $K$. The result is a constant energy decay rate:
\begin{equation}
\partial_\tau\left<\mathcal E(\tau)\right>\approx 2T_{\rm sim}K\alpha_K I(\alpha_K).
\label{eq:scaling}
\end{equation}
Here, $\alpha_K=k\gamma^{-1}\mathcal E_0$ (which depends indirectly on particle number since $\gamma$ must fulfill a fluctuation-dissipation relation) and
\begin{align}
I(z) &= \frac1{2\pi}\int_0^{2\pi}\mathrm{d}\tau\, \sin2\tau\int_{-\infty}^\tau\mathrm{d}\tau'\, e^{-z\int_{\tau'}^\tau\mathrm{d}\tau''\, \cos^2\tau''}\nonumber\\
&\approx -\left(4+\sqrt\pi\left[\tfrac z2\coth\tfrac z2-1\right]\right)^{-1}.
\label{eq:iz}
\end{align}
Eqs.~\eqref{eq:scaling}-\eqref{eq:iz} are valid for both the graphene and carbon nanotube resonators, but the parameter-dependent values of $\TS$ and $\alpha_K$ will change.

\begin{figure}[t]
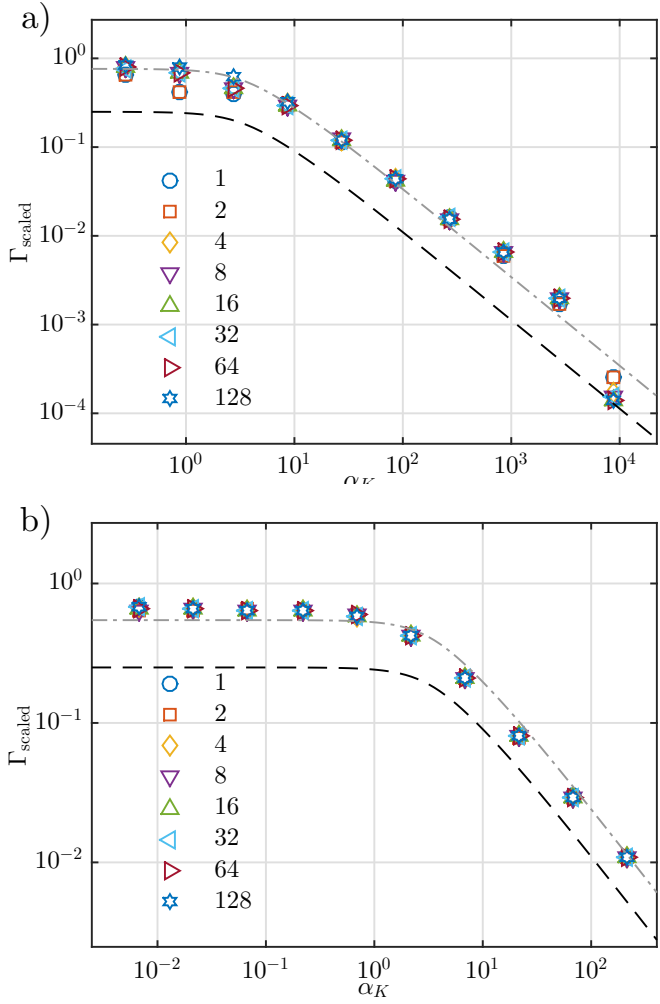

\centering
\includegraphics[width=\linewidth]{figure3a}
\includegraphics[width=\linewidth]{figure3b}
\caption{(color online) Initial dissipation rate $\Gamma_{\rm scaled}$, in the trapping regime, together with $|I(z)|$ (black dashed line), given by equation~\eqref{eq:iz}, as functions of $\alpha_K$ for a) a carbon nanotube resonator and b) a graphene resonator. The number of adsorbates is indicated by the legend. In order to more clearly show the excellent qualitative agreement between theory and simulation, we also plot $|I(z)|$ shifted vertically (gray dash-dotted line) to fit the data points.}
\label{fig:scaling}
\end{figure}

In order to verify the validity of Eq.~\eqref{eq:scaling}, the ringdown of the system was calculated for a broad range of parameter values $K$ and $\alpha_K$, by solving the stochastic differential equations~\eqref{eq:actang1}-\eqref{eq:actang3} using a second-order algorithm~\cite{Mannella_1989, Manella_2000}. The slope $\partial_\tau\mathcal E$ of the initial linear decay of energy was extracted. Simulations were made both while keeping $m$ constant and varying particle number $K$ independently, as well as while keeping the total added mass $m_{\rm tot}=Km$ constant and varying $K$. The result is shown in figure~\ref{fig:scaling}, where $\Gamma_{\rm scaled} = |\partial_\tau\left<\mathcal E\right>/2T_{\rm sim}K\alpha_K|$ as well as $|I(z)|$ are plotted for a wide range of parameter values. The qualitative agreement between the simulation results and the analytical result based on Eqs.~\eqref{eq:scaling} and \eqref{eq:iz} is quite excellent; to illustrate this, Fig.~\ref{fig:scaling} also shows $I(z)$ vertically shifted to fit the data points. The exact vertical shift, $\log 3.05$ in~\ref{fig:scaling}a) and $\log 2.18$ in~\ref{fig:scaling}b), was determined by a least-squares fit.

For large values of $z$, $I(z)\rightarrow -2/z\sqrt\pi$. Interestingly, equation~\eqref{eq:scaling} then reduces to
\begin{equation}
\partial_\tau\left<\mathcal E\right>\approx -\frac4{\sqrt\pi}\TS K=-\frac4{\sqrt\pi}\frac{k_{\rm B}T}{ML^2\omega_0^2}K.
\label{eq:bigz}
\end{equation}
That is, the decay rate is a function of only the temperature and the resonator dimensions: parameters that are generally known a priori to any measurement. Since less-easily determined material constants (e.g. $D$) cancel out, the relationship~\eqref{eq:bigz} is ideally suited for experimental verification of the present work.

Eq.~\eqref{eq:bigz} may seem strange, but can be understood as an upper limit to the single mode-dissipation rate due to particle diffusion. Such a rate is achieved in a situation where, during half a resonator period, a particle moves out from the antinode by gaining energy $\sim k_{\rm B}T$ from thermal fluctuations. This energy is subsequently dissipated by the resonator as it does work on the particle to move it back to the antinode, leading to an average dissipation rate of the order of $\omega_0 k_{\rm B}T$. Each particle adsorbed on the resonator contributes equally to this dissipation of vibrational energy, leading to the prefactor $K$ in~\eqref{eq:bigz}. This process is schematically depicted in Fig.~\ref{fig:dynamics}.

 \subsection{Many adsorbates in the diffusive regime \label{subsec:Diffregime_single_mode}}
 \begin{figure}[t]
\centering
\includegraphics[width=\linewidth]{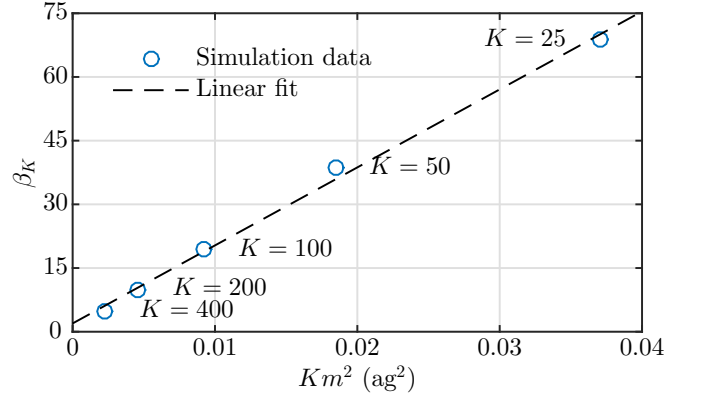}
\caption{(color online) Linearized diffusive regime dissipation rates $\beta_K$ for different numbers of particles $K$, where the total adsorbed mass $Km$ is kept constant. In this regime, the dissipative dynamics of adsorbates causes nonlinear damping of the resonator vibration energy: $\left<{\cal E}(\tau)\right>^{-1}={\cal E}(0)^{-1}+\beta_{K}\tau$. Varying particle number while keeping the total mass $Km$ constant reveals the scaling $\beta_K\propto Km^2$. }
\label{fig:Kscaling}
\end{figure}

When the vibration amplitude is too low for the inertial forces to trap the adsorbates, the particles will instead diffuse freely across the entire surface of the resonator. For a single particle, this leads~\cite{Edblom_2014} to a decay of the average energy akin to that of a nonlinearly damped resonator:
\begin{equation}
\left<{\cal E}(\tau)\right>=\frac{{\cal E}(0)}{1+\beta_1 {\cal E}(0)\tau}.
\label{eq:freedecay}
\end{equation}
The coefficient $\beta_1$ is here given by
\begin{align}
\beta_1&=\epsilon^2\TS^{-1}\sum_n f_n^2 \frac{\lambda_n}{\lambda_n^2+16}\nonumber\\
&=m^2\frac{L^2\omega_0^2}{Mk_{\rm B}T}\sum_n f_n^2 \frac{\lambda_n}{\lambda_n^2+16},
\label{eq:alpha}
\end{align}
with $f_n\equiv \int {\rm d}\xi\, \cos(n\pi\xi)\phi_0(\xi)^2$ and $\lambda_n={\cal D}n^2\pi^2$. 

Generalizing the analysis in Ref.~[\onlinecite{Edblom_2014}] to $K$ particles, we find the same dynamics but with $\beta_K\approx K\beta_1$. From Eq.~\eqref{eq:alpha} it thus follows that $\beta_K\propto Km^2=m_{\mathrm{tot}}m$: both the total adsorbed mass and the individual particle mass affects the rate of dissipation of $\left<\mathcal E\right>$. The fact that the properties of individual adsorbates become relevant in the diffusive regime is not unexpected; to zeroth order, the behavior of the particles is free diffusion, determined by the single-particle diffusion constant $\mathcal D$.

By inverting Eq.~\eqref{eq:freedecay}, the decay rate $\beta_K$ can be found by making a numerical fit to the time evolution of the inverse resonator mode energies. The obtained values of $\beta_K$ are shown in figure~\ref{fig:Kscaling}, which verifies the $Km^2$-scaling for $K\lesssim10^2$.
\begin{figure}[t]
\centering
\includegraphics[width=\linewidth]{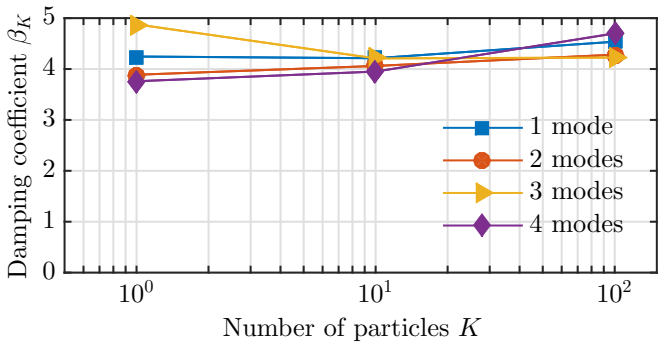}
\caption{(color online)  Linearized decay rates $\beta_K$ in the diffusive regime for different numbers of particles $K$, with $Km^2$ constant. The lines are a guide to the eye, labeling different numbers of flexural modes included in the calculation. Whereas the number of modes included in the trapping regime does affect the decay rate, there is no significant dependence on the number of modes in the diffusive regime.}
\label{fig:Modenumberscaling}
\end{figure}

\subsection{Multi-mode dynamics in the diffusive
regime\label{subsec:Diffregime_many_modes}}

Finally, we turn to the matter of including multiple flexural modes in the simulations. In Ref.~[\onlinecite{Edblom_2014}], it was found that for a single particle, the decay rate in the trapping regime increases as more modes are included in the calculations. This is because the decay rate limit of $\sim \omega_0 k_{\rm B}T$ can be overcome, since energy can be transferred not only to the adsorbate, but also to the other modes through the diffusion-induced mode coupling. We have already seen that for many particles, the scaling with particle number is less obvious in the diffusive regime, with the dissipation scaling as $\Gamma\sim Km^2$. Here, we consider how the dissipation rate in the diffusive regime changes as the number of flexural modes are increased.

The full equations of motion~\eqref{eq:eom1}-\eqref{eq:eom2} were numerically integrated, and the vibrational energy evolution was fitted to the expression~\eqref{eq:freedecay}. The resulting value for the decay constant $\beta_K$ in the diffusive regime is shown in Fig.~\ref{fig:Modenumberscaling}. In the simulations, \mbox{$Km^2\approx 0.002$~ag$^2$} was held fixed while the number of particles $K$ and the number of flexural modes included in the calculation were both
varied. The numbers used correspond to a single-walled carbon nanotube resonator with a length of 1~$\rm{\mu}$m and diameter of $5$~nm, having a fundamental mode frequency $\omega_0=2\pi\times361$~MHz at $T=300$~K. The diffusion constant was chosen as $D=0.1$~cm$^2$/s. 

As can be seen from Fig.~\ref{fig:Modenumberscaling}, including more modes in the calculations in the diffusive regime does not affect the ring-down time; the multimode dynamics are both qualitatively and quantitatively equivalent to the single-mode dynamics.

\section{Conclusions}
We have the studied diffusion-induced dissipation in carbon nanotube and graphene resonators, in a situation where $K$ non-interacting particles, each with mass $m$, are adsorbed on the resonator surface. We find that a linear scaling of the fundamental mode energy decay rate with the total adsorbed mass, $\Gamma\propto Km= m_{\rm tot}$, holds in the inertial trapping regime. In the diffusive regime we instead find a scaling $\Gamma\propto Km^2=m_{\rm tot}m$. Additionally, in contrast to the dynamics in the trapping regime, the intermode coupling responsible for the equilibration of the higher flexural modes does not significantly affect the ring-down dynamics of the fundamental mode in the diffusive regime.  

In Ref.~[\onlinecite{Edblom_2014}], diffusion-induced dissipation was discussed in the context of mass sensing, where a single adsorbate is purposefully deposited on a resonator. However, the dissipation mechanism will certainly be present also in the case of particles being unintentionally adsorbed. As we here have showed, for several particles the resulting energy decay rates can exceed the single-particle ones by orders of magnitude; it follows that our results are highly relevant for, e.g., experiments performed under ambient conditions. While it is necessary that the diffusion constants are relatively high, we note that diffusion on, for instance, graphene can be enhanced by thermal ripples, as recently shown in Ref.~[\onlinecite{Ma_2015}].

\begin{acknowledgments}
We acknowledge financial support from the Swedish Research Council VR (AI), the Foundation for Strategic  Research SSF (CR), and the European Union through grant no. 604391 (CR). 
\end{acknowledgments}




\end{document}